\documentclass[final,5p,times,twocolumn]{elsarticle}
\usepackage{graphicx}
\usepackage{amsmath, amsthm, amssymb}
\journal{Ecological Complexity}
\begin{document}
\begin{frontmatter}
\title{Spatio-temporal stochastic resonance induces patterns in wetland vegetation dynamics}

\author[poli]{S. Scarsoglio}
\author[UV]{P. D'Odorico}
\author[poli]{F. Laio}
\author[poli]{L. Ridolfi}

\address[poli]{Department of Water Engineering, Politecnico di Torino, Torino, Italy}
\address[UV]{Department of Environmental Sciences, University of Virginia, Charlottesville, Virginia, USA}

\begin{abstract}

Water availability is a major environmental driver affecting
riparian and wetland vegetation. The interaction between water
table fluctuations and vegetation in a stochastic environment
contributes to the complexity of the dynamics of these ecosystems.
We investigate the possible emergence of spatial patterns induced by spatio-temporal stochastic resonance in a
simple model of groundwater-dependent ecosystems. These
spatio-temporal dynamics are driven by the combined effect of
three components: (i) an additive white Gaussian noise, accounting
for external random disturbances such as fires or fluctuations in
rain water availability, (ii) a weak periodic modulation in time,
describing hydrological drivers such as seasonal fluctuations of
 water table depth, and (iii) a spatial coupling term, which
takes into account the  ability of vegetation to spread and colonize other parts of the landscape. A
suitable cooperation between these three terms is able to give
rise to ordered structures which show spatial and temporal
coherence, and are statistically steady in time.

\end{abstract}


\end{frontmatter}

\section{Introduction}

Random fluctuations can play an  important role in determining the
composition and structure of ecosystems \citep{Borgogno2009,Holling1973}. Fire occurrences \citep{D_Odorico_2007},
fluctuations in rain water availability \citep{Noy_Meir_1973,Laio_2001,D_Odorico_2006}, soil heterogeneity
\citep{Puigdefabregas1999}, and temperature oscillations
\citep{Polyak1996}, are some of the examples of fluctuating
environmental drivers. Evaluating ecosystem response to
environmental fluctuations is crucial, among other things, to the
management and restoration of plant communities \citep{Wright2002,Ridolfi2006}, the understanding of their
ability to sustain biodiversity \citep{Naiman1997,Mitsch2000,D_Odorico2008}, and to predict
causes of possible shifts in  vegetation composition \citep{Ridolfi2007,Chambers2001,Wright2002}.

The effects of noise on dynamical systems has been studied in a
number of stochastic models proposed in recent years \citep{Sagues2007,GarciaOjalvo1999}. Noise is typically
associated with a "destroying effect", as a driver of disorganized fluctuations around the stable states of the
underlying deterministic dynamics. However, a number of studies
have demonstrated the existence of a "constructive effect" of noise:
 ordered spatial structures
(i.e. patterns) can emerge in some spatio-temporal dynamical
systems as a result of noisy fluctuations (see, for example,
\citep{Ridolfi_book_2011,Scarsoglio2011,VanDenBroeck97,Parrondo1996,Sieber2007}). An increase in noise intensity  can produce in these
systems a counterintuitive, more regular and organized, behaviour
both in time and in space.

For example, the cooperation of a random external  driver with an
external periodic forcing is able to induce ordered transitions
between different states of a dynamical system. Known as
stochastic resonance (see, among others, \citep{Gammaitoni1998,Lindner1995,Vilar1997}), this phenomenon may
occur when a bistable system is disturbed by an external random
forcing in the presence of a weak periodic fluctuation in time.
Bistable system exhibit a bistable potential with two minima
separated by a potential barrier. The periodic forcing causes
fluctuations in the height of the potential barrier, without ever
causing a transition between the two states. It is the external
random forcing - if strong enough - that is able to drive the
system from a stable state to the other; these transitions are
more likely to occur when the height of the potential barrier is
lowest, which depends on the phase of the periodic fluctuations.
 A suitable synchronization between the
frequency of the random transitions (which is related to the noise
intensity) and the frequency of the periodic forcing induces a
 resonance-like effect that generates regular transitions between the
two stable states \citep{Wellens2004,Perc2008}.

Even though a number of studies have recognized the
 bistable character  of some ecosystems \citep{Rietkerk1997,Brovkin1998,Zeng2004,Ridolfi2006} and the ability of random fluctuations to
induce new dynamical behaviours that do not exist in the
underlying deterministic dynamics \citep{Borgogno2007,D_Odorico2005}, applications of stochastic resonance theories to
eco-hydrology have started to appear only recently (see, for
instance, \citep{Borgogno2011,Spagnolo2004,Rao2009,Sun2010}).

In this paper we study the basic mechanisms for the occurrence of
spatio-temporal stochastic resonance in a simple model describing
the dynamics of riparian or wetland vegetation dynamics. The
temporal deterministic dynamics of the model have been proposed by
Ridolfi et al. \citep{Ridolfi2006} to investigate the emergence of bistability
from the interactions between phreatophyte vegetation and shallow
phreatic aquifers in wetland ecosystems. The combined effect of
external noise and periodic oscillations on these
bistable dynamics has been shown to be able to induce stochastic
resonance in the time domain \citep{Borgogno2011}.

We investigate the possible emergence of spatio-temporal
stochastic resonance when a diffusive spatial coupling term is
introduced in the presence of random and periodic forcings. The addition of a spatial diffusive mechanism accounts
for the vegetation ability to encroach across the landscape. The other
assumptions are physically based, due to the presence of environmental disturbances (e.g.
occurrence of fires, rainfall fluctuations) acting on the system, and a number of periodic
hydrological drivers (e.g., seasonal oscillations of the water
table). In particular, a significant connection exists between water table fluctuations and vegetation dynamics. The water level variation appears to be an independent and important vegetation
gradient, as the responses of species to the range of water level fluctuations
seem to reflect their tolerance to disturbances \citep{Laitinen2008}. Changes in species composition and distribution, as well as in vegetation
structure, have been related to various factors, mainly the vegetation topographical position
with respect to the water table, frequency and duration of
inundation \citep{Moreno1999}. It has been observed that, under the periodic declining groundwater availability, vegetation patches tend to show a regular spatial behaviour \citep{Kong2009}. Moreover, in dune systems, the occurrence of regular patterns of heath patches has been associated, among other things, to the different groundwater discharge \citep{Munoz2005}.

The relevant question here is whether the phenomenon of stochastic
resonance - resulting from the combined effect of noise and a weak
temporal oscillation - can interact with a spatial diffusive term
to generate vegetation ordered spatial structures.


\section{Modeling framework}

We consider the dynamics of plant biomass, $V$, and its interplay
with the phreatic aquifer \citep{Ridolfi2006}. We first recall
some important aspects of the temporal deterministic dynamics, and
then present the characteristics of the spatio-temporal stochastic
system.
\subsection{Deterministic model}
Changes in vegetation biomass are the result of a growth-death
process that can be expressed as (e.g., \citep{Noy_Meir_1975,Tsoularis2002}),

\begin{equation}
\frac{\textmd{d}V}{\textmd{d} \tau}= V (V_{cc} - V),
\label{deterministic_model}
\end{equation}

\noindent where $V$ is the dimensionless biomass, normalized  by a
(high) fixed reference value, $\tau=\alpha t$ is the dimensionless
time (where $\alpha$ controls the temporal response of the system and
$t$ is time), and $V_{cc}$ is the dimensionless ecosystem carrying
capacity, that is the maximum  amount of vegetation sustainable
with the available resources.

The interaction between phreatophyte vegetation, i.e. plants
relying on the phreatic aquifer, and the average depth of the
local water table is widely recognized as one of the key aspects
affecting wetland ecosystems dynamics \citep{Peck1987,Roy2000,Chang2002,Ridolfi2006}. For example,
the water table decreases in the presence of phreatophyte species
because of the lower recharge rates due to rainfall interception
and plant transpiration \citep{Wilde1953,Borg1988,Riekerk1989,Dube1995}, and also because vegetation
taproots directly extract water from the aquifer \citep{Lemaitre1999}.
The depth, $d$, of the water table is adimensionalized with
respect to  water table depth in the absence of vegetation, $d^*$,
and can be expressed as a linear function of plant biomass, $V$,

\begin{equation}
d = 1 + \beta V,
\label{d_deterministic}
\end{equation}

\noindent where $\beta$ is the dimensionless sensitivity of  the
water table to the presence of vegetation. $\beta$ is positive
because plants typically tend to increase the depth of the water
table. The water table depth, $d$, in turn affects the dynamics of
wetland vegetation. If the water table is too shallow
(waterlogging), vegetation can suffer, due to an insufficient
aeration of the root zone and a decreased rate of seedling
establishment \citep{Roy2000}. If the water table is too deep,
water is out of reach of taproots and vegetation can suffer as
well. These effects are taken into account by a quadratic
dependence of the carrying capacity, $V_{cc}$, on the water table
depth, $d$,

\begin{equation}
\centering
V_{cc}=
\begin{cases}
a (d - d_{inf}) (d_{sup} - d) \,\,\,\,\,\,\,\, & \mbox{if }d_{inf} < d < d_{sup}, \\
0  \,\,\,\,\,\,\,\, & \mbox{otherwise},
\end{cases}
\label{V_cc_deterministic}
\end{equation}

\noindent where $a$ regulates the  sensitivity of carrying
capacity to changes in water table depth, $d_{inf}$ and $d_{sup}$
are the adimensional thresholds of vegetation tolerance to shallow
and deep water tables, respectively.

A positive feedback may exist between vegetation establishment and
water table dynamics, whereby phreatophytes favor their own
survival by increasing the water table depth and enhancing root
aeration \citep{Wilde1953,Chang2002}. The presence of
positive feedback mechanisms suggests the possible existence of
multiple stable states \citep{Ridolfi2006}. This fact is
confirmed by some experimental evidence suggesting that two
alternative equilibrium states may exist in wetland vegetation
dynamics \citep{Roy2000,Chambers2001,Wright2002,Schroder2005}. Equilibrium states can
be found by setting Eq. (\ref{deterministic_model}) equal to zero.
$V=V_0=0$ is always an equilibrium state, therefore the existence
of multiple equilibrium states depends on the existence of real,
nonnull roots of the equation $V_{cc}=V$. These solutions can be
found by substituting Eq. (\ref{d_deterministic}) in Eq.
(\ref{V_cc_deterministic}), and then searching for intersections
between $V_{cc}(V)$ and $V_{cc}=V$ (see Fig. \ref{V_cc_V}). Ridolfi et al. \citep{Ridolfi2006} showed three possible cases:

\begin{figure}
\centering
\begin{minipage}[]{0.32\columnwidth}
   \includegraphics[width=\columnwidth]{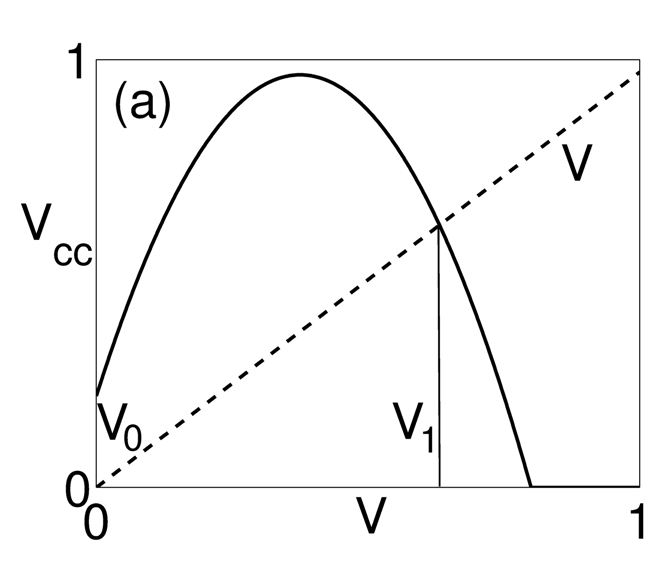}
\end{minipage}
\begin{minipage}[]{0.32\columnwidth}
   \includegraphics[width=\columnwidth]{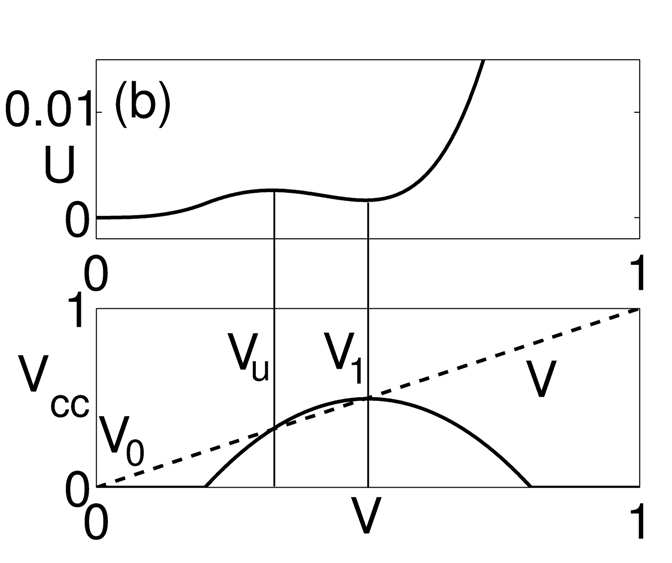}
\end{minipage}
\begin{minipage}[]{0.32\columnwidth}
   \includegraphics[width=\columnwidth]{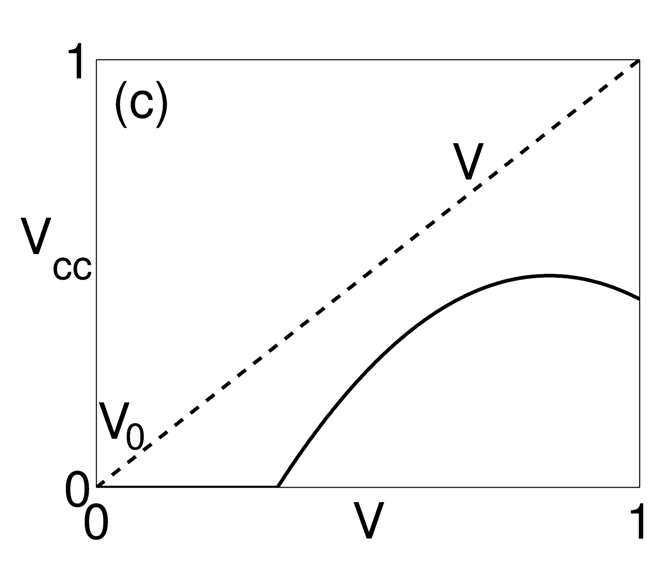}
\end{minipage}
\caption{Vegetation carrying capacity $V_{cc}$ as
function of the vegetation biomass $V$, with $a=11$ and $d_{sup}=1.8$. (a) $\beta=1 $, $d_{inf}=0.95$.
(b) $\beta=1$, $d_{inf}=1.2$. In the upper panel the potential
$U(V)$ is shown. (c) $\beta=0.6$, $d_{inf}=1.2$.} \label{V_cc_V}
\end{figure}

\noindent (a) $d_{inf} < 1$ (Fig. \ref{V_cc_V}a). The water table
depth in the absence of vegetation, $d=1$, is always greater than
the minimum depth required for vegetation establishment. $V_0=0$
is an unstable state and only one stable state exists, $V_1$.
Ecosystem dynamics always tend the vegetated state, $V_1$,
regardless the initial conditions.

\noindent (b) $d_{inf} > 1$ (Fig. \ref{V_cc_V}b) and vegetation is
able  to keep the water table below the maximum elevation
acceptable for plants (e.g. $\beta=1$), three equilibrium states
are found. Two of these states, $V=V_0$ and $V=V_1$ are stable,
while the other one, $V=V_u$, is unstable. If $V$ grows above
$V_u$, then the system tends to the vegetated state. Otherwise, it
evolves towards the unvegetated state and remains here blocked.
Note that the two stable states correspond to the minima of the
double-well potential, U(V), with
$\frac{\textmd{d}V}{\textmd{d}\tau}=-\frac{\textmd{d}U}{\textmd{d}V}$,
while the unstable solution, $V=V_u$ corresponds to a maximum
 or "potential barrier".

\noindent (c) $d_{inf} > 1$ (Fig. \ref{V_cc_V}c) and vegetation is
not able to reduce the water table elevation below the minimum
level suitable for plant survival (e.g. $\beta=0.6$). In this
case  the only possible stable state is the unvegetated condition,
$V=V_0=0$.

\subsection{Spatio-temporal stochastic model}

We now focus on the case (b) described above, with two stable
states (vegetated and unvegetated) emerging from the deterministic
dynamics. We introduce in these dynamics  three new terms: (i) a
periodic temporally oscillating term, accounting for seasonal or interannual
fluctuations of the water table depth, $d$, \citep{Rosenberry1997,Zhang2006}; (ii) a random forcing
term, representing the effect of random environmental drivers
(e.g., fires or climate fluctuations); (iii) a spatial coupling
term, modeling the diffusive spread of vegetation.

\noindent Eq. (\ref{deterministic_model}) now reads,

\begin{equation}
\frac{\partial V}{\partial \tau}= V (V_{cc} - V) + \xi({\bf r},\tau) + D \nabla^2 V
\label{stochastic_model}
\end{equation}

\noindent where $D$ is the strength of the spatial coupling,
$\nabla^2$ is the Laplace operator, $\xi({\bf r},\tau)$ is a white
(in time and space) Gaussian noise with zero mean, intensity $s$,
and correlation $\langle \xi({\bf r},\tau) \xi({\bf r'},\tau')
\rangle = 2 s \delta({\bf r}-{\bf r'})\delta(\tau-\tau')$, being $\delta(\cdot)$ the Dirac delta function. The
temporal oscillation is directly inserted into Eq.
(\ref{stochastic_model}), through the dependence of $V_{cc}$ on
the water table depth, Eq. (\ref{d_deterministic}), which can be
now expressed as

\begin{equation}
d = 1 + A \textmd{cos} \left(\frac{\omega \tau}{\alpha}\right) + \beta V,
\label{d_stochastic}
\end{equation}

\noindent where $A$ and $\omega=\frac{2 \pi}{T}$ are the
nondimensional amplitude  and the dimensional frequency of the
seasonal oscillations, respectively. $T$ is the oscillation period
(for example $T=$ 1 year). Notice that these seasonal fluctuations of the water table are crucial to the emergence of the dynamical behaviors (i.e., spatiotemporal stochastic resonance) presented and discussed in the following sections. Therefore, Eq.
(\ref{V_cc_deterministic}) now reads

\begin{equation}
\centering
V_{cc}=
\begin{cases}
a \left[1 + A \textmd{cos} \left(\frac{\omega \tau}{\alpha}\right) + \beta V - d_{inf}\right]\left[d_{sup} - 1 - A \textmd{cos} \left(\frac{\omega \tau}{\alpha}\right) - \beta V\right] \\
 \,\,\,\,\,\,\,\,\,\,\,\,\,\,\,\,\,\,\,\,\,\,\,\, \mbox{if }d_{inf} < d < d_{sup}, \\
0 \,\,\,\,\,\,\,\,\,\,\,\,\,\,\,\,\,\,\,\,\,\,\,\, \mbox{otherwise}.
\end{cases}
\label{V_cc_stochastic}
\end{equation}

\begin{figure}
\centering
   \includegraphics[width=0.5\columnwidth]{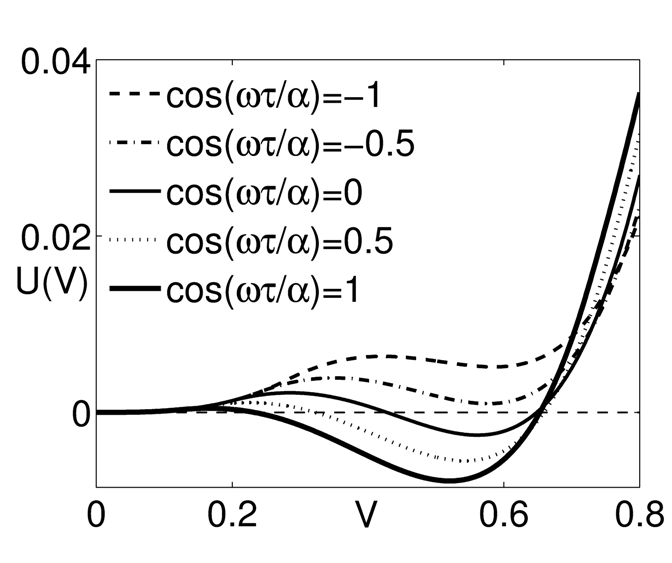}
\caption{Potential $U(V)$ of the temporal  deterministic model
(\ref{deterministic_model}) at different times during the periodic cycle of $d$ (\ref{d_stochastic}). $A=0.08$,
$\alpha=0.5$ d$^{-1}$, $\omega=0.0344$ rad/d, $a=13$, $d_{inf}=1.2$,
$d_{sup}=1.8$, $\beta=1$.} \label{potential}
\end{figure}

We consider the case of relatively weak seasonal oscillations in a
way that the potential, $U(V)$,  still has two minima
corresponding to two stable states. Fig. \ref{potential} shows the
potential, $U(V)$, of the temporal deterministic model
(\ref{deterministic_model}), but with the water table depth, $d$,
modulated as in (\ref{d_stochastic}). The potential always
maintains two minima throughout its periodic oscillation. Thus,
even though there are phases in the modulation of the periodic
forcing in which the height of the potential barrier is strongly
reduced (hence, the probability of occurrence of a state
transition is enhanced), deterministic transitions between the two
wells of the potential are not allowed.

The choice of an additive white (Gaussian) noise to
represent external drivers acting on vegetation dynamics in the time domain is
motivated by the fact that the temporal scales of the
random driver are typically much shorter than the characteristic temporal
scales over which the temporal vegetation dynamics evolve
(e.g. the relaxation time to a stable state). Moreover we used a white noise also in space to show how spatial coherence is not imposed by the spatial correlation of these external driver but emerges as an effect of spatio-temporal stochastic resonance. Therefore, this kind
of noise is typically adopted in stochastic modeling \citep{Sagues2007}.

The Laplacian, $\nabla^2$, in Eq.
(\ref{stochastic_model}) is a simple operator which is widely used
to represent the spatial effects of the diffusion mechanisms in
vegetation dynamics \citep{Borgogno2009,Manor2008,vonHardenberg2010}. This operator accounts for spatial interactions
between a point of the domain and its nearest neighbors, and is
therefore considered as a short-range spatial coupling.

\section{Results} \label{Results}

\begin{figure}
\centering
\begin{minipage}[]{0.32\columnwidth}
   \includegraphics[width=\columnwidth]{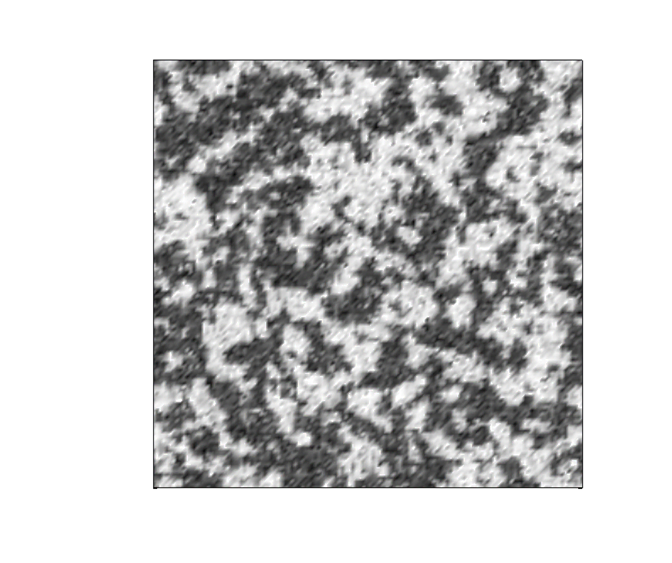}
\end{minipage}
\begin{minipage}[]{0.32\columnwidth}
   \includegraphics[width=\columnwidth]{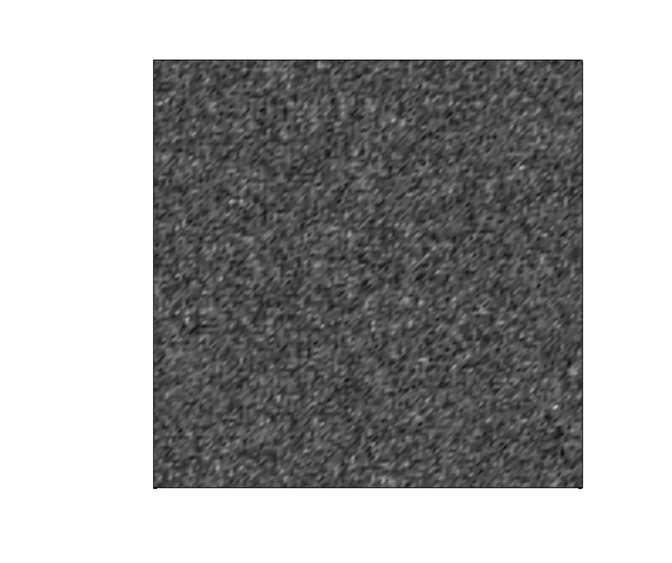}
\end{minipage}
\begin{minipage}[]{0.32\columnwidth}
   \includegraphics[width=\columnwidth]{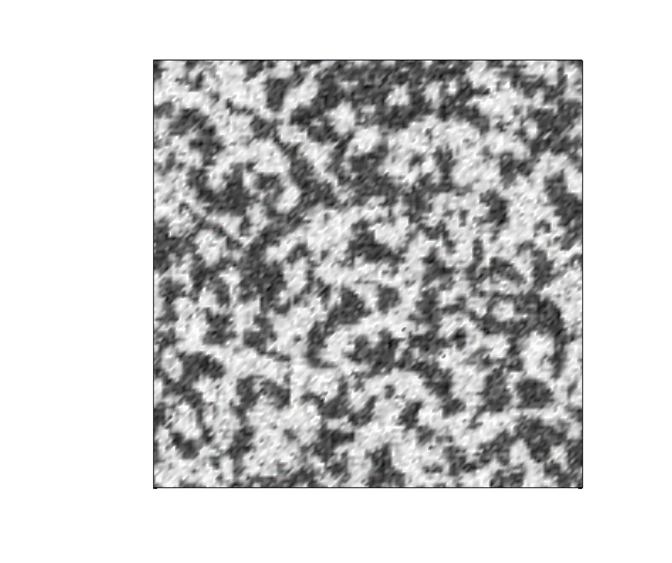}
\end{minipage}
\begin{minipage}[]{0.32\columnwidth}
   \includegraphics[width=\columnwidth]{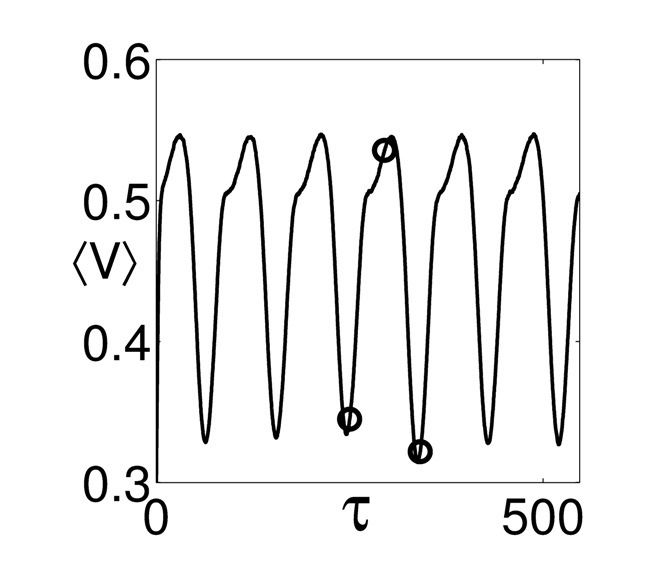}
\end{minipage}
\begin{minipage}[]{0.32\columnwidth}
   \includegraphics[width=\columnwidth]{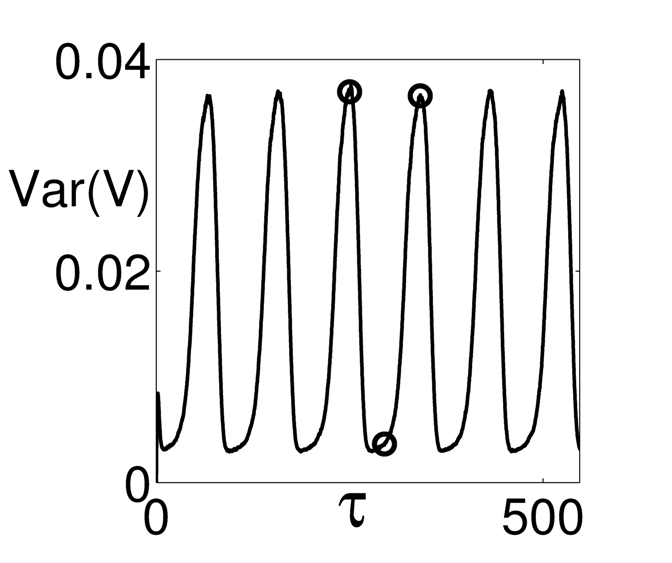}
\end{minipage}
\begin{minipage}[]{0.32\columnwidth}
   \includegraphics[width=\columnwidth]{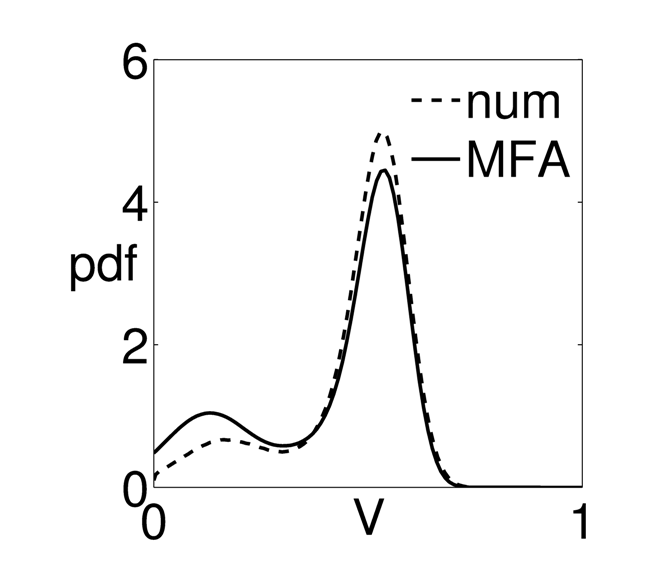}
\end{minipage}
\caption{Model (\ref{stochastic_model}), $s=0.008$, $D=0.1$, $A=0.08$, $\alpha=0.5$ d$^{-1}$, $\beta=1$, $a=13$, $d_{sup}=1.8$, $d_{inf}=1.2$, $T=182.5$ days. Top: numerical simulations at $\tau=250, 295, 341$. Bottom: mean and variance of the vegetation biomass, $V$, as functions of time (black circles correspond to mean and variance values at $\tau=250, 295, 341$). Pdf of vegetation field (solid: mean-field analysis, dashed: numerical evaluation).}
\label{results_s_0_002_D_0_05}
\end{figure}

The model (\ref{stochastic_model}) is numerically solved for
different parameter sets to directly investigate the associated
spatio-temporal vegetation dynamics. Periodic boundary conditions are
adopted, while, in the absence of other indications, the initial
conditions used in the simulations are uniformly distributed
random numbers in the range [0.29, 0.31]. We also use the mean field
analysis, a mathematical framework that provides the steady-state
probability distribution of $V$ \citep{Ridolfi_book_2011,Vandenbroeck1994,VanDenBroeck97}. Details about the numerical scheme and the mean-field analysis are reported
in \ref{num_ana}.

Fig. \ref{results_s_0_002_D_0_05} shows numerical results for
$s=0.008$ and $D=0.1$ (other parameter values are reported in the
caption of the figure). Black and white tones are used for the
highest ($V=0.7$) and lowest ($V=0$) values of $V$, respectively.
Well defined patterns periodically emerge (e.g., at $\tau=250$ and
$\tau=341$) and disappear ($\tau=295$) in the course of
periodic fluctuations of $d$. In this case the non-dimensional period is
$T'=91.25$. Pattern occurrence corresponds to temporal minima
 of mean vegetation biomass and to maxima of biomass variance, while
homogeneous vegetated states correspond to maxima of mean biomass
and minima of the variance of vegetation biomass (see circles in
Fig. \ref{results_s_0_002_D_0_05}). Here and in the following
discussion, we call "homogeneous" a state in which the system
exhibits negligible spatial heterogeneity with respect to the
patterned state.

The spatio-temporal dynamics continuously oscillate
between two differently vegetated states. In this example patterns appear
during the transition between these two states, when the system
approaches the least vegetated stable state, and disappear when
the most vegetated state is reached. The interaction of noise with
periodic water table oscillations and spatial diffusion gives rise to spatial ordered structures which are statistically
steady and periodically appear/disappear in time with the same
period, $T$, as the fluctuations in $d$. These periodic state
transitions, and the associated formation and
disappearance of spatial patterns are the evidence of
spatio-temporal stochastic resonance, with noise-induced shifts
between the two minima of the potential, $U(V)$. Although the
system tends to oscillate between two stable states, these states
are not always reached. Indeed, temporal oscillations have to be
sufficiently long to permit the system to visit both states. Here,
for example, the temporal period, $T=182.5$ days, is too short and
does not allow the system to reach the unvegetated state.
Conversely, with a longer period, $T$, the system is able to
attain both the vegetated and the unvegetated states (see Section
\ref{different_periodicity}).

In Fig. \ref{results_s_0_002_D_0_05} numerical and analytical evaluations of the pdf of the vegetation field are reported. Here and in the following discussion, the numerical approximation of the pdf (dashed curves) is obtained using simulated fields in the temporal range $50<\tau<547$, which encompasses several temporal oscillation periods (here, $T'=91.25$). Details on the analytical approximation of the pdf (solid curves) are discussed in \ref{num_ana}. The agreement between numerical and analytical evaluations of the pdf is quite good in showing a weak bimodality.


\subsection{Role of $D$ and $s$}

\begin{figure}
\centering
\begin{minipage}[]{0.32\columnwidth}
   \includegraphics[width=\columnwidth]{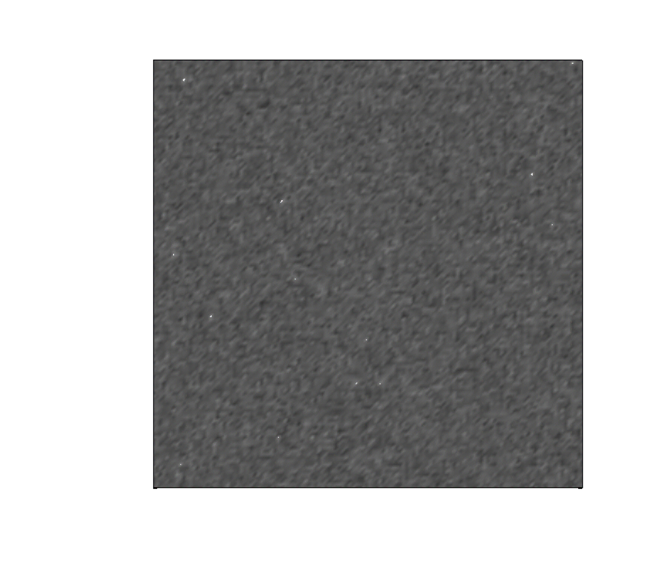}
\end{minipage}
\begin{minipage}[]{0.32\columnwidth}
   \includegraphics[width=\columnwidth]{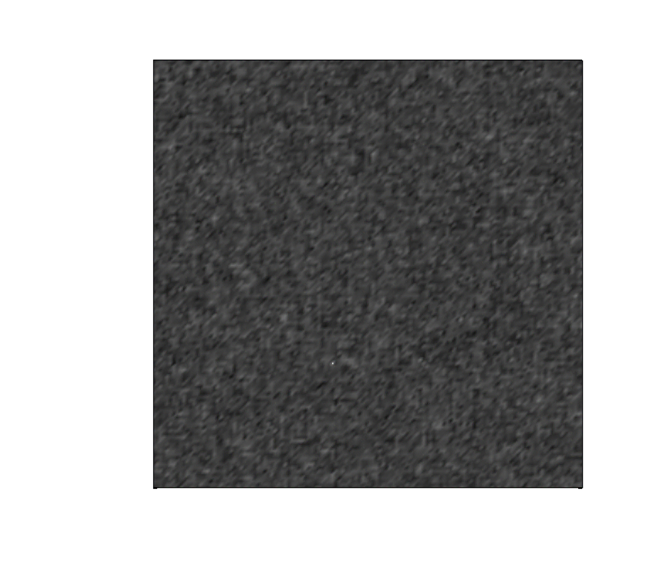}
\end{minipage}
\begin{minipage}[]{0.32\columnwidth}
   \includegraphics[width=\columnwidth]{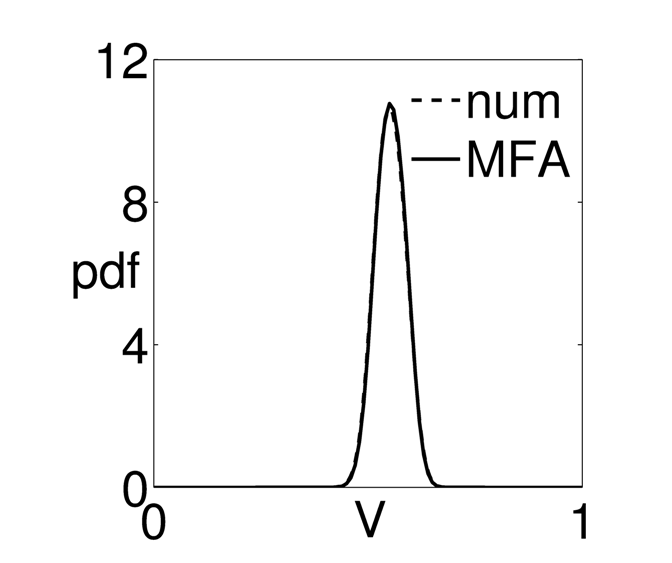}
\end{minipage}
\begin{minipage}[]{0.32\columnwidth}
   \includegraphics[width=\columnwidth]{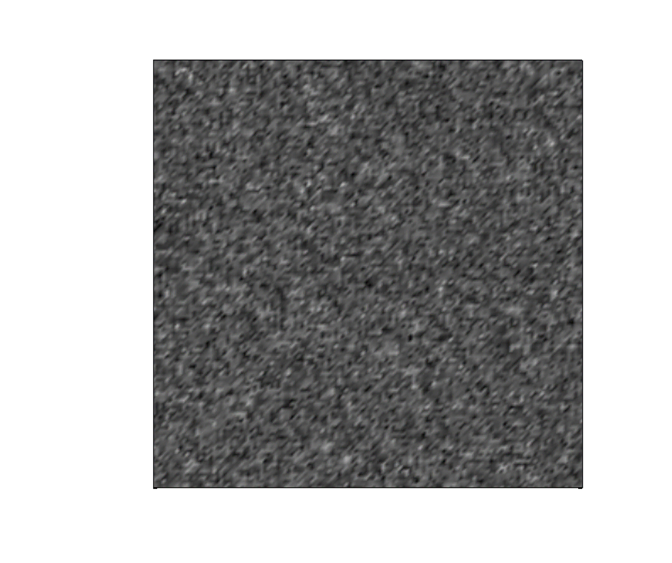}
\end{minipage}
\begin{minipage}[]{0.32\columnwidth}
   \includegraphics[width=\columnwidth]{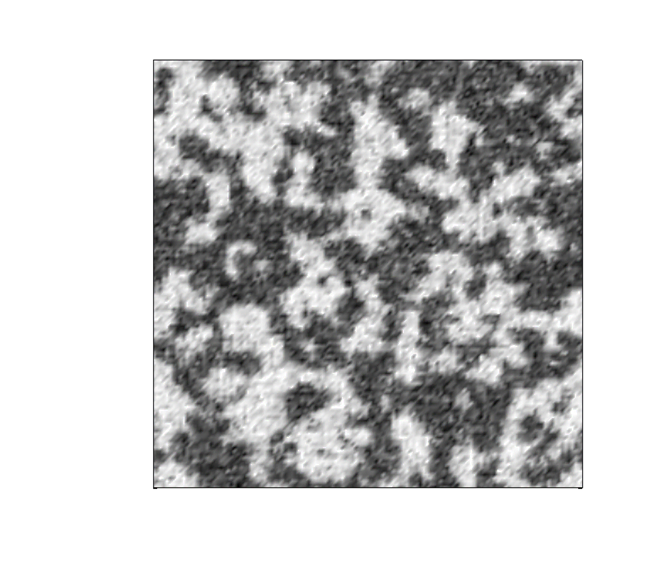}
\end{minipage}
\begin{minipage}[]{0.32\columnwidth}
   \includegraphics[width=\columnwidth]{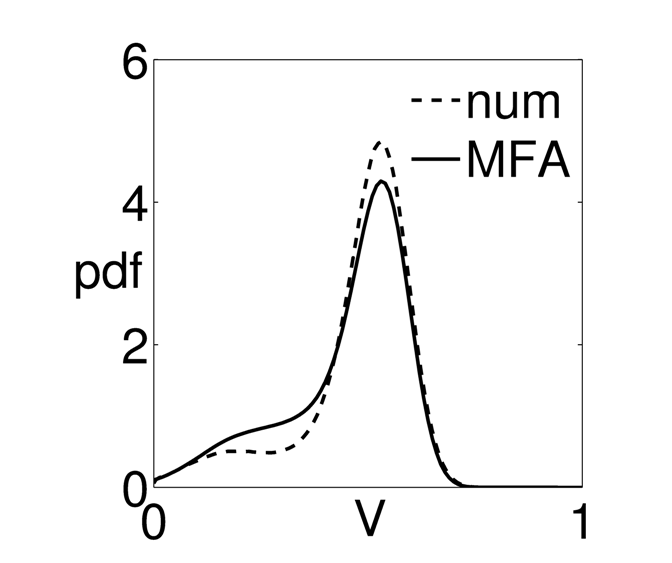}
\end{minipage}
\begin{minipage}[]{0.32\columnwidth}
   \includegraphics[width=\columnwidth]{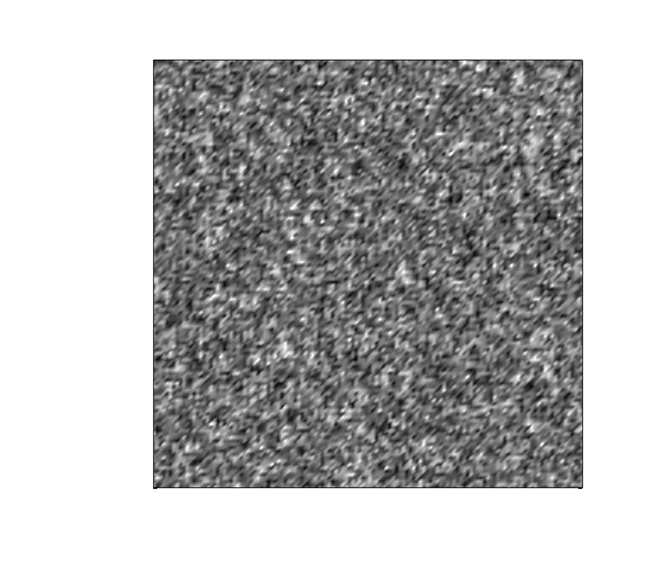}
\end{minipage}
\begin{minipage}[]{0.32\columnwidth}
   \includegraphics[width=\columnwidth]{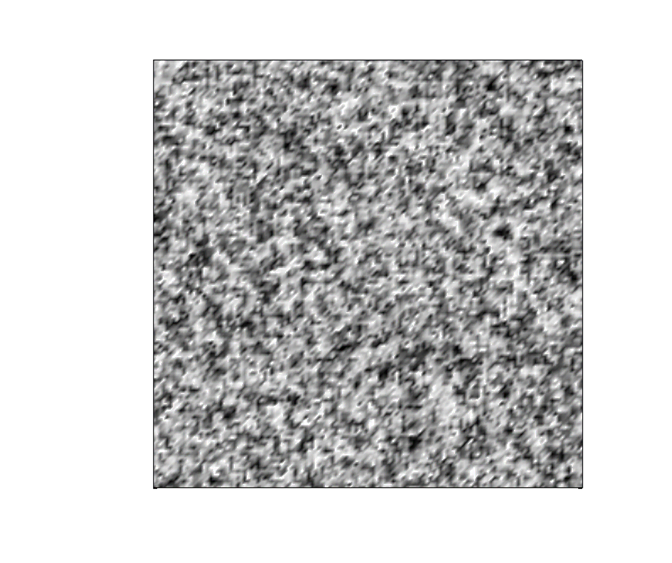}
\end{minipage}
\begin{minipage}[]{0.32\columnwidth}
   \includegraphics[width=\columnwidth]{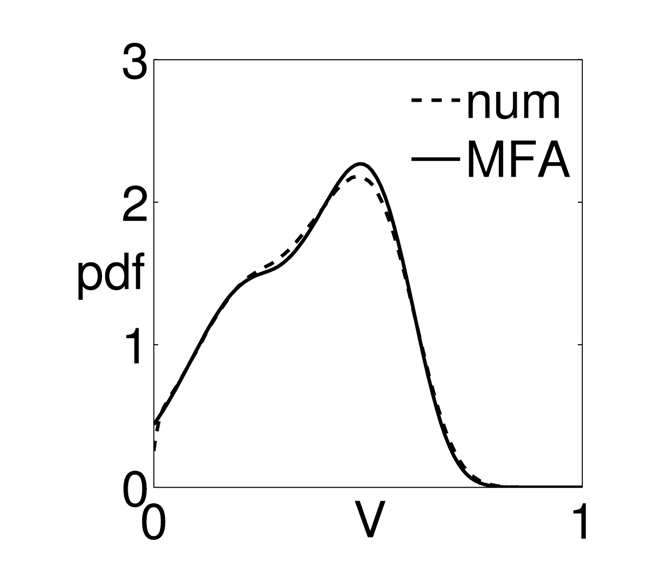}
\end{minipage}
\caption{Numerical simulations of model (\ref{stochastic_model}) with $D=0.2$, other parameters as in Fig.
\ref{results_s_0_002_D_0_05}. (Top row) Field distributions with $s=0.0024$ at $\tau=275$ (left column),
$\tau=320$ (middle column), and pdf (right column). (Middle row) Field distributions with $s=0.012$ at $\tau=202$ (left column),
$\tau=247$ (middle column), and pdf (right column). (Bottom row) Field distributions with $s=0.04$ at $\tau=270$ (left column), $\tau=310$ (middle column),
and pdf (right column).} \label{results_D_0_1_s_0_0006_0_003_0_01}
\end{figure}

We now consider changes in the spatial coupling strength, $D$ ($D=0.01,
0.1, 1$), with a fixed noise intensity (e.g. $s=0.008$). As $D$
increases, the least vegetated state moves closer the most
vegetated one. As a consequence, for higher $D$ values, the system
becomes more homogeneous (the biomass variance oscillates  around
lower values and the amplitude of these fluctuations becomes
smaller). By increasing $D$ from $0.01$ to $0.1$, spatially
coherent patterns appear. If $D$ is further increased, the diffusive effect is
so strong that the field is almost homogeneous, and the system
experiences a blocking effect induced by diffusion.

Different noise levels ($s=0.0024, 0.012, 0.04$) with a fixed
spatial coupling ($D=0.2$) are then considered. If the noise
intensity is weak ($s=0.0024$) with respect to the spatial
coupling strength, $D$, the biomass density tends to be
homogeneous in time and space. On the other hand, if the noise
level is too strong ($s=0.04$), the system experiences random
oscillations between more vegetated and less vegetated states.
However, in this case noise has a "destroying" effect in that it
destroys any spatial and temporal coherence and no ordered states
emerge. An intermediate noise value ($s=0.012$) allows for the
emergence of temporal fluctuations  and the formation of periodic
spatial patterns. The existence of an intermediate noise level
which, in cooperation with a weak temporal oscillation, is able to
regularly move the system between different states is the most
typical feature of the stochastic resonance phenomenon \citep{Wellens2004,Gammaitoni1998}. In Fig.
\ref{results_D_0_1_s_0_0006_0_003_0_01}, we show the configuration
of the system in correspondence to local minima ($\tau=275$ for $s=0.0024$, $\tau=202$ for $s=0.012$, $\tau=270$ for $s=0.04$) and maxima ($\tau=320$ for $s=0.0024$, $\tau=247$ for $s=0.012$, $\tau=310$ for $s=0.04$) of
biomass variance (left and central columns, respectively) for the
three values of noise intensity.

In general, temporal local maxima (minima) of biomass variance
 correspond to minima (maxima) of biomass mean only when
the system is able to exhibit ordered spatial structures (see Fig.
\ref{results_s_0_002_D_0_05}). In all the other cases -- i.e.
with homogeneous (Fig. \ref{results_D_0_1_s_0_0006_0_003_0_01},
top) or disturbed (Fig. \ref{results_D_0_1_s_0_0006_0_003_0_01},
bottom) spatial distributions -- there is not such a
synchronization between maxima (minima) of biomass variance and
minima (maxima) of biomass mean.

Also in these cases, there is a very good agreement between the mean-field
steady-state pdfs (solid curves) and the pdfs numerically
calculated (dashed curves) using simulated fields in the
temporal range $50<\tau<547$. This result is especially
interesting because this mean-field analysis accounts for the
 modulation of the dynamics in time (see \ref{num_ana}), an approach that has never been reported before.

From the results here presented, we can conclude that
there is an optimal region in the $(s, D)$ plane, in which spatial
patterns emerge. This region corresponds to a noise intensity,
$s$, of order $10^{-2}$, and  a spatial coupling strength, $D$, of
order of magnitude $10^{-1}$.

\subsection{Influence of initial conditions}
\begin{figure}
\centering
\begin{minipage}[]{0.32\columnwidth}
   \includegraphics[width=\columnwidth]{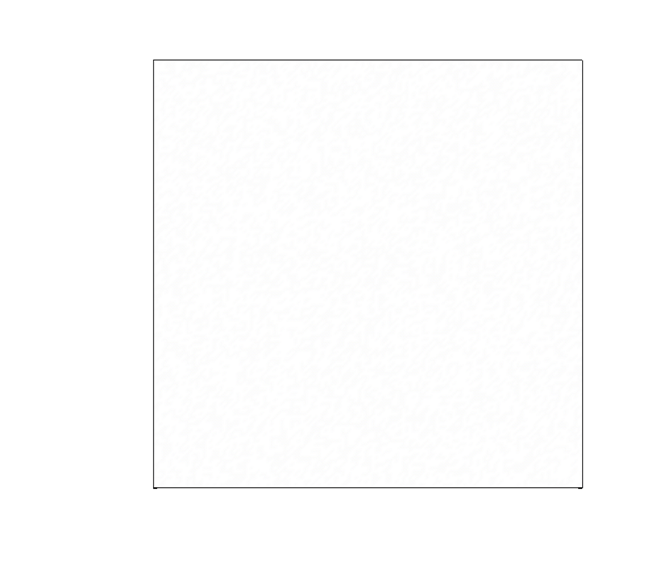}
\end{minipage}
\begin{minipage}[]{0.32\columnwidth}
   \includegraphics[width=\columnwidth]{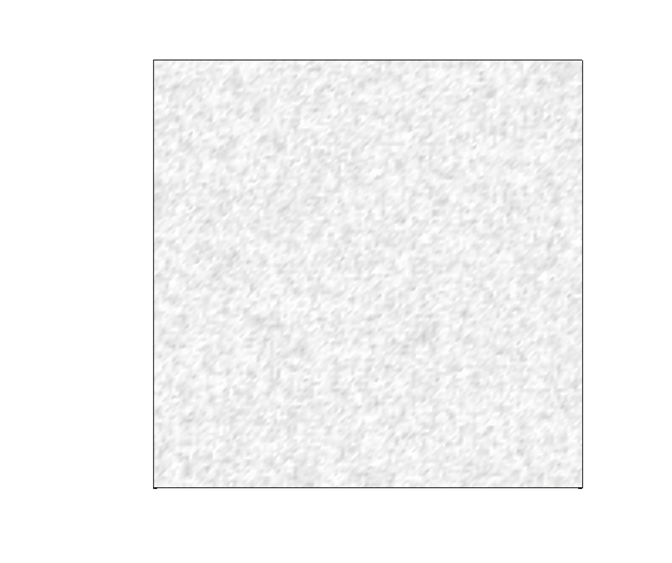}
\end{minipage}
\begin{minipage}[]{0.32\columnwidth}
   \includegraphics[width=\columnwidth]{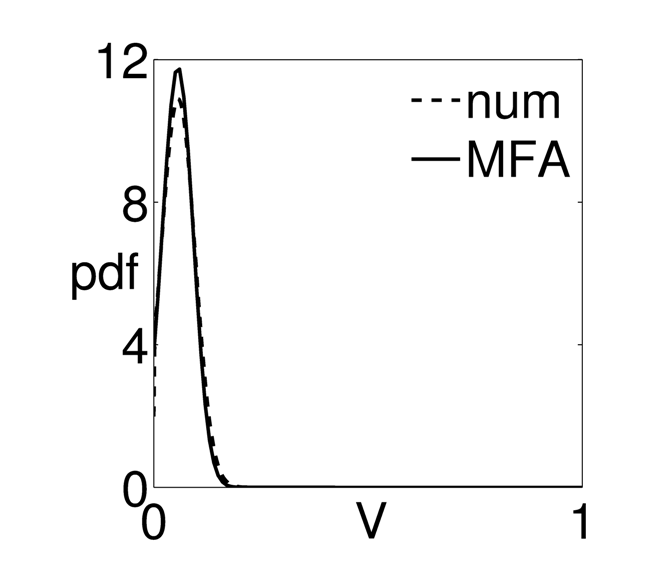}
\end{minipage}
\begin{minipage}[]{0.32\columnwidth}
   \includegraphics[width=\columnwidth]{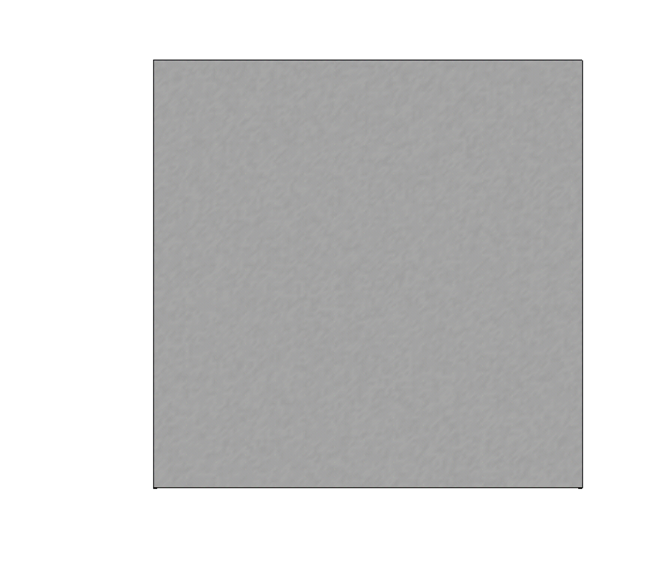}
\end{minipage}
\begin{minipage}[]{0.32\columnwidth}
   \includegraphics[width=\columnwidth]{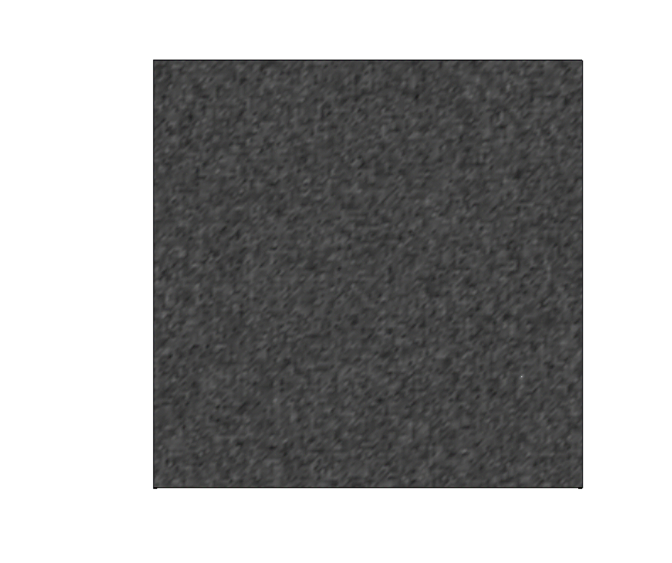}
\end{minipage}
\begin{minipage}[]{0.32\columnwidth}
   \includegraphics[width=\columnwidth]{pdf_s_0_0006_D_0_1.png}
\end{minipage}
\caption{Model (\ref{stochastic_model}), $s=0.0024$, $D=0.2$, other parameters as in Fig. \ref{results_s_0_002_D_0_05}. (Top row): Field distributions with random initial conditions uniformly distributed in the range [0, 0.02] at $\tau=0$ (left column), $\tau=250$ (middle column), and pdf (right column). (Bottom row): Field distributions with random initial conditions uniformly distributed in the range [0.29, 0.31] at $\tau=0$ (left column), $\tau=250$ (middle column), and pdf (right column).}
\label{initial_conditions}
\end{figure}

In the previous sections the initial conditions were generated as
random numbers uniformly distributed in the interval [0.29, 0.31]. If noise is
sufficiently strong with respect to the spatial coupling strength,
$D$, results similar to those shown in figures
\ref{results_s_0_002_D_0_05}, and
\ref{results_D_0_1_s_0_0006_0_003_0_01} (middle and bottom), are
obtained with different initial conditions. Conversely, if the noise intensity, $s$, is small with
respect to $D$, the system tends to an homogeneous steady state,
which depends on the initial conditions. In other words, a sufficiently high noise intensity with respect to the spatial coupling strength allows the system to continuously oscillate in time between the two minima of the potential, regardless the initial conditions. On the contrary, if noise is small with respect to the spatial coupling strength, the system remains trapped into the minimum of the potential which is closer to the initial conditions.

An example is shown in
Fig. \ref{initial_conditions} where, for $s=0.0024$ and $D=0.2$,
different initial conditions, uniformly distributed in the
interval [0, 0.02] or [0.29, 0.31], lead to an unvegetated (top)
or a vegetated (bottom) state, respectively. When the initial
conditions are in the interval [0, 0.02], the system is initially
close to the minimum of the potential ($V_0=0$). The noise
intensity $s$ is so weak with respect to the spatial coupling
strength, $D$, that it is not able to move the system away from
this stable state; therefore the system remains blocked into the
unvegetated stable state.

\subsection{Different periodicity of the temporal modulation}\label{different_periodicity}

\begin{figure}
\centering
\begin{minipage}[]{0.32\columnwidth}
   \includegraphics[width=\columnwidth]{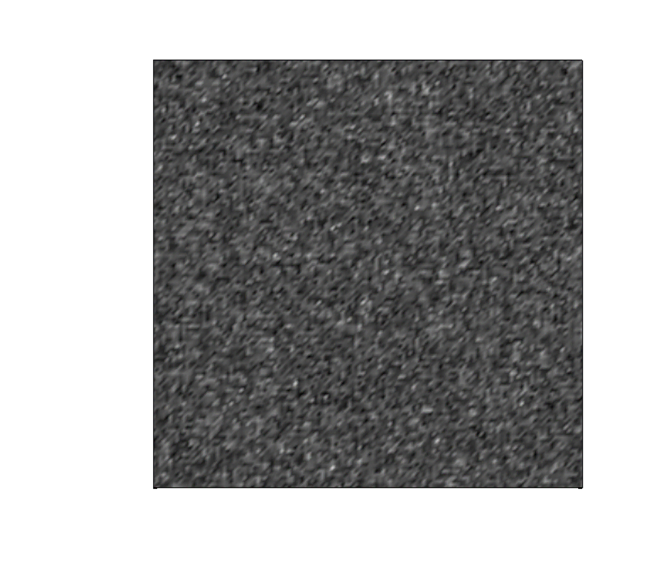}
\end{minipage}
\begin{minipage}[]{0.32\columnwidth}
   \includegraphics[width=\columnwidth]{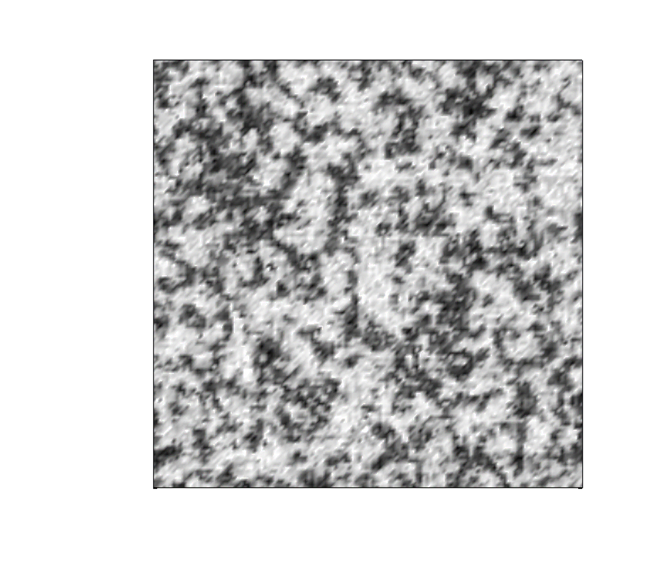}
\end{minipage}
\begin{minipage}[]{0.32\columnwidth}
   \includegraphics[width=\columnwidth]{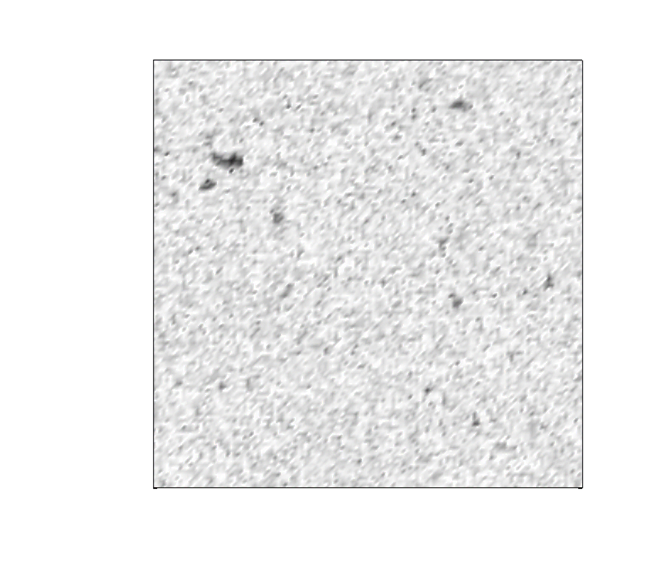}
\end{minipage}
\begin{minipage}[]{0.32\columnwidth}
   \includegraphics[width=\columnwidth]{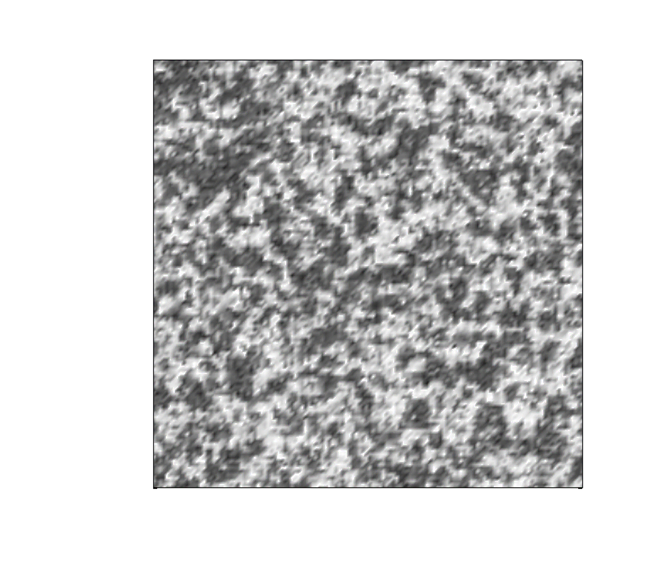}
\end{minipage}
\begin{minipage}[]{0.32\columnwidth}
   \includegraphics[width=\columnwidth]{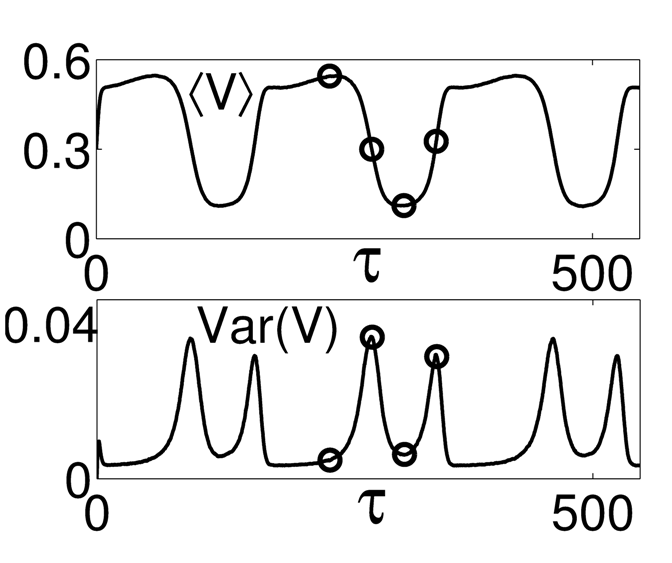}
\end{minipage}
\begin{minipage}[]{0.32\columnwidth}
   \includegraphics[width=\columnwidth]{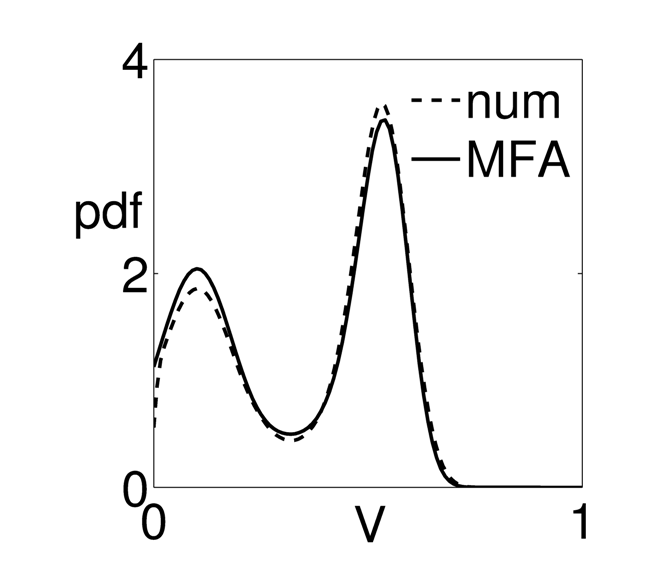}
\end{minipage}
\caption{Model (\ref{stochastic_model}), $s=0.008$, $D=0.1$, $A=0.08$, $\alpha=0.5$ d$^{-1}$, $\beta=1$, $a=13$, $d_{sup}=1.8$, $d_{inf}=1.2$, $T=365$ days. Numerical simulations at $\tau=235, 277, 310, 342$. Bottom: mean and variance of the vegetation biomass, $V$, as functions of time (black circles correspond to mean and variance values at $\tau=235, 277, 310, 342$). Pdf of vegetation field (solid: mean-field analysis, dashed: numerical evaluation).}
\label{results_s_0_002_D_0_05_T_365}
\end{figure}

We then evaluate the effect of a different temporal periodicity in the modulation
term, $G = \textmd{cos}\left(\frac{\omega \tau}{\alpha}\right)$,
by taking $T=365$ days, while all the other parameters remain the same as in Fig. \ref{results_s_0_002_D_0_05}.
Patterns periodically emerge ($\tau=277$ and $\tau=342$) and disappear ($\tau=235$ and $\tau=310$).
Their occurrence corresponds to temporal local maxima of the vegetation variance, while homogeneous vegetated
and unvegetated stable states correspond to maxima and minima of the vegetation mean,
respectively (see circles in Fig. \ref{results_s_0_002_D_0_05_T_365}). A longer period, $T$, allows the system to
explore the whole range of values of $V$. This fact is confirmed by a non-negligible bimodality of both analytical and numerical pdfs, whose peaks correspond to the two alternatively visited stable states. During these state transitions, patterns appear. Although the temporal modulation establishes the frequency of pattern appearance, the mechanism of pattern formation is always present and does not depend on the specific temporal periodicity imposed into the system. 

\subsection{Non-simultaneous presence of random and periodic forcings}
\begin{figure}
\centering
\begin{minipage}[]{0.22\columnwidth}
   \includegraphics[width=\columnwidth]{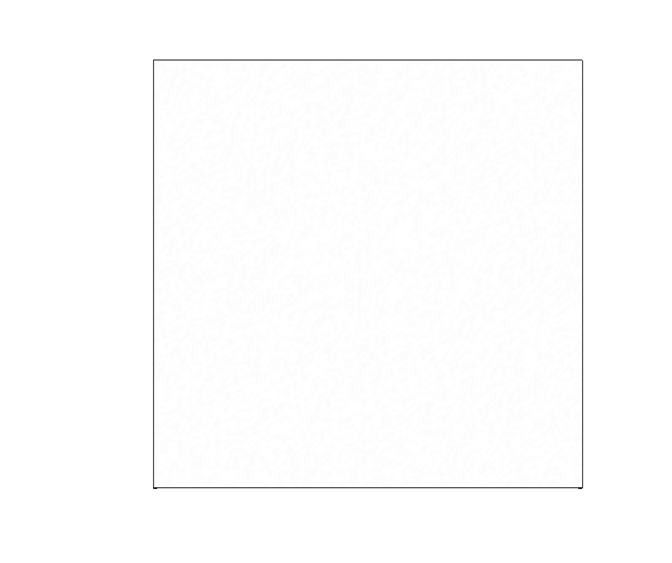}
\end{minipage}
\begin{minipage}[]{0.22\columnwidth}
   \includegraphics[width=\columnwidth]{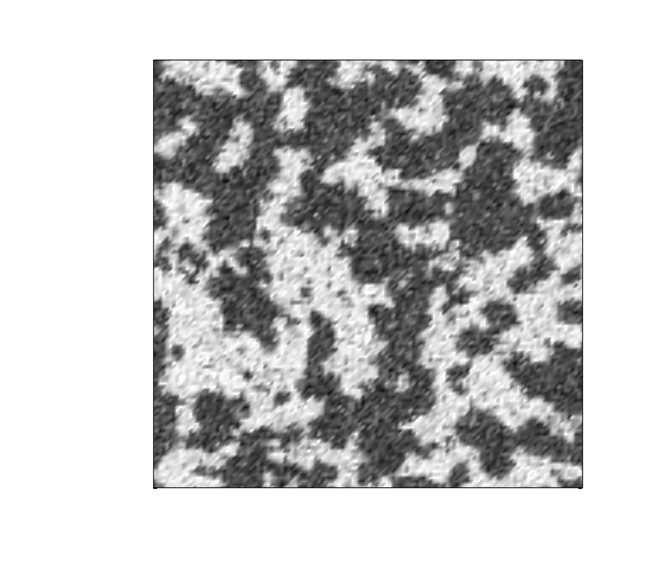}
\end{minipage}
\begin{minipage}[]{0.22\columnwidth}
   \includegraphics[width=\columnwidth]{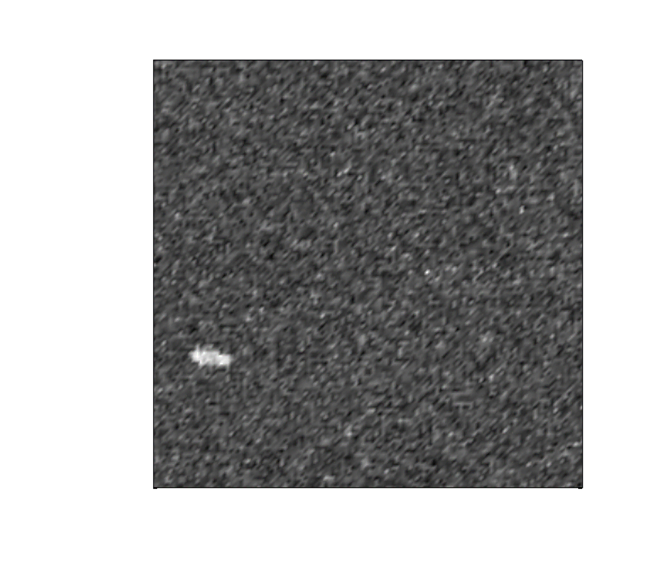}
\end{minipage}
\begin{minipage}[]{0.22\columnwidth}
   \includegraphics[width=\columnwidth]{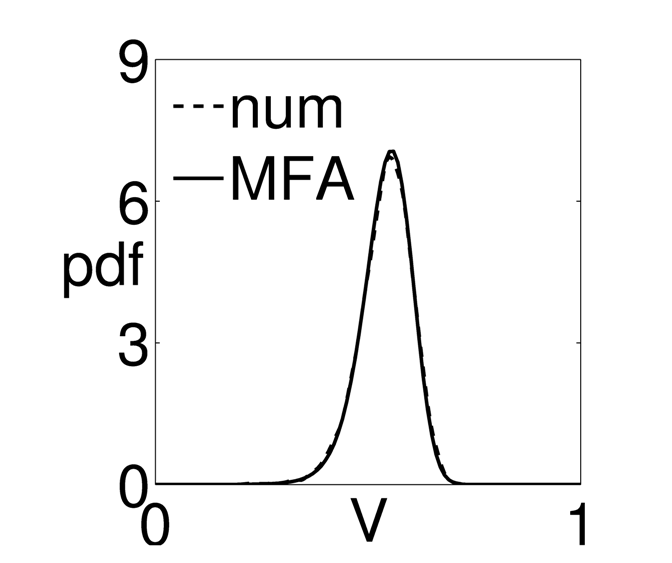}
\end{minipage}
\begin{minipage}[]{0.22\columnwidth}
   \includegraphics[width=\columnwidth]{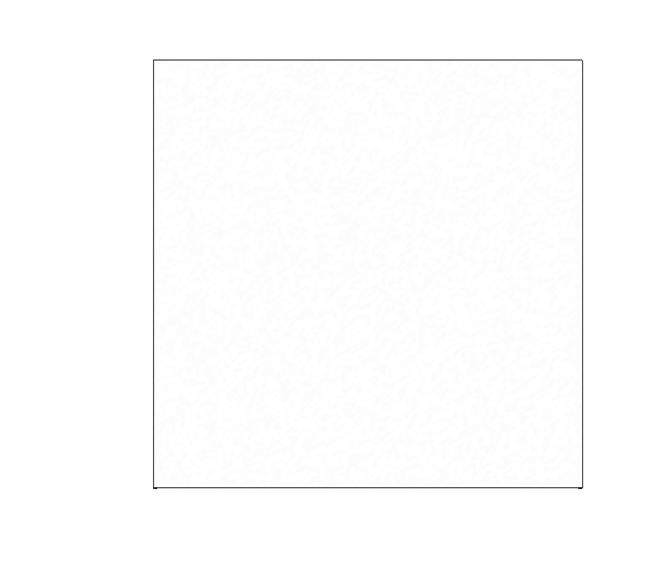}
\end{minipage}
\begin{minipage}[]{0.22\columnwidth}
   \includegraphics[width=\columnwidth]{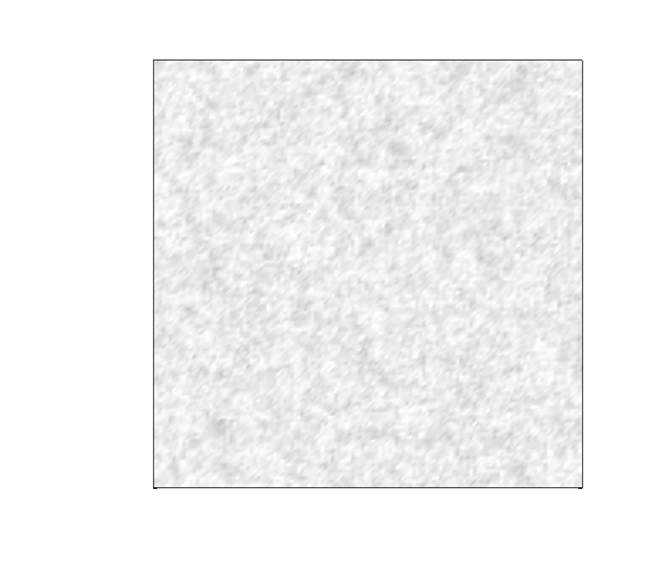}
\end{minipage}
\begin{minipage}[]{0.22\columnwidth}
   \includegraphics[width=\columnwidth]{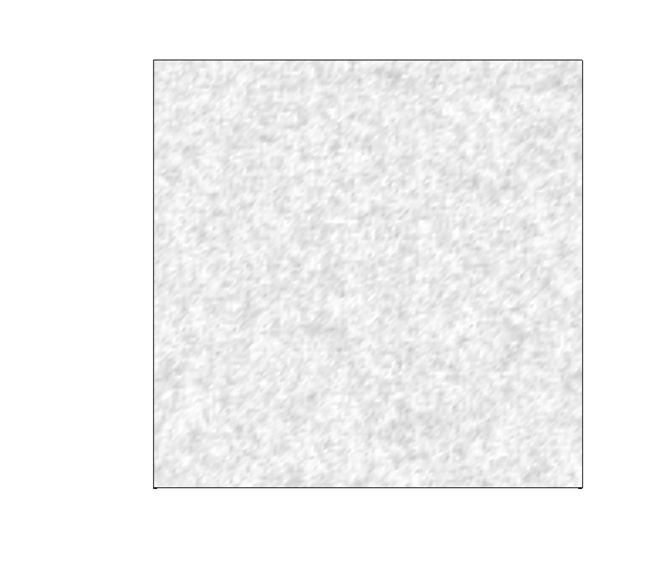}
\end{minipage}
\begin{minipage}[]{0.22\columnwidth}
   \includegraphics[width=\columnwidth]{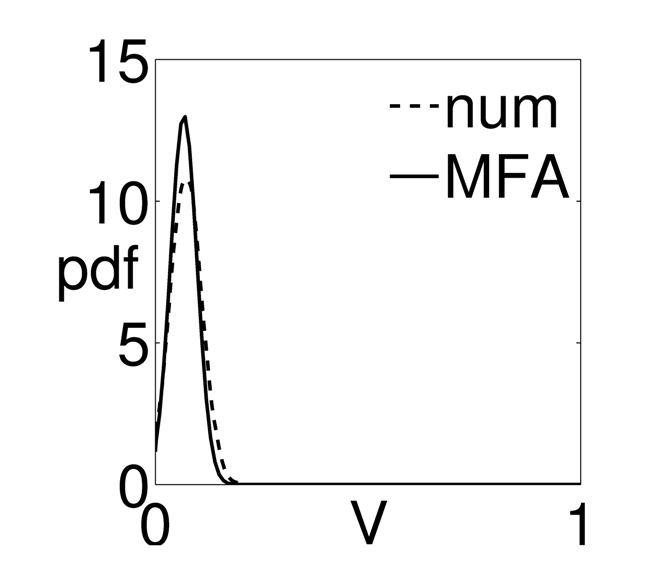}
\end{minipage}
\caption{Model (\ref{stochastic_model}) with $s=0.008$, $A=0$, other parameters as in Fig. \ref{results_s_0_002_D_0_05}, initial conditions are random numbers uniformly distributed in the interval [0, 0.02]. (Top row) Field distributions with $D=0.1$ at $\tau=0, 125, 250$ (from left to right), and pdf (right column, dashed: numerical at $\tau=547$, solid: mean-field analysis). (Bottom row) Field distributions with $D=1$ at $\tau=0, 125, 250$ (from left to right), and pdf (right column, dashed: numerical at $\tau=547$, solid: mean-field analysis).}
\label{results_s_0_002_D_0_05_0_5_A_0}
\end{figure}

We now consider the case in which noise and temporal modulation do
not occur at the same time. As noted, as the noise level, $s$,
decreases the spatial distribution of plant biomass becomes more
homogeneous. A similar behavior is expected to occur for $s=0$.
The homogeneous state depends on the initial conditions: if the
system is initially in the interval [0, 0.02] it homogeneously
decays to zero, while for initial conditions in the interval
[0.29, 0.31] it reaches the homogeneous vegetated stable state and
weakly oscillates in time around it.

\noindent When noise is present with no modulation ($A=0$), state
transitions may occur. If the noise is weak with respect to the
strength of the spatial coupling, $D$, the homogeneous steady
state attained by the system depends on the initial conditions. If
the noise is strong enough with respect to $D$, the system tends
to  the homogeneous vegetated state regardless of the initial
conditions. Patterns can  emerge only as transient features
before the system approaches the homogeneous steady state. An
example is shown in Fig. \ref{results_s_0_002_D_0_05_0_5_A_0}
where $s=0.008$, the initial conditions are in the interval [0,
0.02], $D=0.1$ (top) and $D=1$ (bottom). For $D=0.1$, while
passing from an unvegetated initial condition ($\tau=0$) to a
homogeneous vegetated steady state ($\tau=250$), the system
exhibits some transient patterns ($\tau=125$), which fade out in
the long run. Instead, for $D=1$ the noise is weak with
respect to $D$ and therefore the system is blocked into the
low vegetated state without even showing transient patterns.

\section{Discussion and conclusions}
In the system investigated in this paper the occurrence of
spatio-temporal stochastic resonance can be thought of as the
result of the combined effect of three components: a white (in
time and space) Gaussian noise, a weak periodic modulation in
time, and a suitable spatial coupling term. In the absence of
modulation, the noise - if strong enough with respect to the
spatial coupling - can lead the system towards the vegetated
state, regardless of the initial conditions. However, if the noise
is weak with respect to $D$, the system will approach one of the
two (vegetated or unvegetated) stable states, depending on the
initial conditions. With no modulation, patterns can only occur
in an initial transient because, once the stable state is reached, the system
is locked in a homogeneous state. On the other hand, in the
absence of noise the temporal modulation  is not able to promote a
state transition, because the system reaches one of the two stable
states -- depending on initial conditions -- and weakly oscillates
around it. The spatial coupling term is able to induce spatial
coherence without itself enhancing state transitions. In other
words, noise is the driver able to induce state transitions, while
the periodic forcing and the spatial coupling are able to induce
temporal and spatial coherence, respectively. Only a suitable
cooperation between these three mechanisms can lead to the
interesting ordered scenarios analyzed in Section \ref{Results}.
Thus, random and periodic forcings have to be simultaneously
present along with a suitable  spatial coupling to induce the
formation of statistically steady ordered structures with spatial
and temporal coherence.

Changes in the periodicity, $T$, of the weak temporal modulation
strongly affect vegetation dynamics. In this case, if $T$ is of
the order of six months or one year, coherent spatial structures
 periodically appear during transitions between vegetated
and unvegetated states. These dynamics could represent the
emergence and disappearance of an herbaceous species, which
typically have a growth/life period comparable to the
driving seasonal oscillations of the water table. Conversely, the
emergence of  patterns in model simulations with longer periods
may represent the formation of ordered structures in the
distribution of tree species, which grow over longer time scales.

\noindent Beside this interpretation based on a single species,
spatial patterns result from the simultaneous existence of two
different species, whose growth/life periods are comparable to $T$
but not synchronized  with one anther (i.e. when species 1 is at
maximum density, species 2 is at minimum density). The first
species is dominant at a certain time (almost homogeneous field,
see, for example, left top panel of Fig.
\ref{results_s_0_002_D_0_05_T_365}), but it tends to disappear,
while the density of other increases. The simultaneous presence of
both may give rise to ordered spatial structures (spatial pattern
occurrence, see middle top panel of Fig.
\ref{results_s_0_002_D_0_05_T_365}). At some point, the second
species temporarily dominates the plant community composition
(almost homogeneous field, see right top panel of Fig.
\ref{results_s_0_002_D_0_05_T_365}), but it will then start to
decay, while the first species grows back and new patterns emerge
(see left bottom panel of Fig.
\ref{results_s_0_002_D_0_05_T_365}).

\appendix
\section{Numerical and analytical methods} \label{num_ana}

The typical numerical approach used to solve stochastic partial
differential equations is based on a discretization of the
continuous spatial domain using a regular Cartesian lattice with
spacing $\Delta x=\Delta y=\Delta$. A two-dimensional square
lattice with 128x128 sites and $\Delta=1$ is here adopted. The
original equation (\ref{stochastic_model}) is then transformed
into a system of coupled stochastic ordinary differential
equations with a finite difference scheme,

\begin{equation}
\label{syst_mf} \frac{{\rm d} V_{i}}{{\rm d} \tau}= V_{i} (V_{cc,i} - V_{i})+
\xi_{i} + D \sum_{j \in nn(i)} (V_j - V_i),
\end{equation}

\noindent where $V_i$ and $\xi_i$ are the values of $V$ and $\xi$
at site $i$,  respectively, $nn(i)$ is the set of the four nearest
neighbors of the site $i$. Numerical simulations
are carried out with the Heun's predictor-corrector scheme \citep{VanDenBroeck97,Sagues2007}, with a temporal step
$\Delta \tau = 5\cdot10^{-5}$. Notice how this method does not prevent the emergence of a spurious spatial correlation associated with the discrete representation of the dynamics in a 2D lattice \citep{Lythe2001}. In our analyses the simulations were repeated with grids of different sizes and the same qualitative patterns and dynamical behaviors were observed to emerge regardless of the resolution used in the discrete representation of the domain.

A qualitative representation of (stochastic) spatio-temporal dynamics
can be obtained through the mean-field method. Its fundamental
assumption is that the mean of the values of $V$ in all the
neighboring cells can be approximated by the spatio-temporal mean
of the field, namely $\sum_{j} (V_j) = 4 \langle V_j \rangle = 4
\langle V \rangle$. Therefore, eq. (\ref{syst_mf}) becomes

\begin{equation}
\label{model_mfa} \frac{{\rm d} V_{i}}{{\rm d} \tau}= V_{i} (V_{cc,i} - V_{i})+
\xi_{i} + 4 D (\langle V \rangle - V_i).
\end{equation}

The classic mean-field analysis allows one to evaluate if a phase
transition occurs. The existence of multiple solutions, $\langle V \rangle_n$, of the self-consistency equation

\begin{equation}\label{self_consistency_m}
\langle V \rangle_n=\int_{0}^{1} V \ p_n^{st}(V|G; \langle V \rangle_n) \ {\rm
d}V=F\bigl(\langle V \rangle_n \bigr),
\end{equation}

\noindent indicates the
presence of a non-equilibrium phase transition. In Eq. (\ref{self_consistency_m}) $G = \textmd{cos}\left(\frac{\omega
\tau}{\alpha}\right)$ is the temporal modulation and
$p_n^{st}(V|G; \langle V \rangle_n)$ is the conditional
probability with respect to $G$ at steady state. We recall that the
occurrence of non-equilibrium phase transition is neither a
necessary nor a sufficient condition for noise-induced pattern
formation \citep{Ridolfi_book_2011,Scarsoglio2011}. In fact,
non-equilibrium phase transitions imply that noise is able to
change the value of the order parameter, but not that ordered
geometrical structures necessarily emerge.

\noindent Here, the dynamics exhibit also a  temporal
modulation, $G$,  which has to be adequately accounted for while
applying the mean-field method. To evaluate the steady state pdf
of the biomass vegetation, $V$, and at the same time to deal with
the temporal modulation $G$, we propose to solve the
self-consistency condition (\ref{self_consistency_m}) for all the
 values of $G \in [-1, 1]$. In this way the conditional
probability, $p_n^{st}(V|G; \langle V \rangle_n)$, is obtained.
The steady-state probability $p^{st}(V; \langle V \rangle)$ is
then computed through the convolution product between the
conditional probability, $p_n^{st}(V|G; \langle V \rangle_n)$, and
the probability, $p(G)$, of the temporal modulation, $G$.

Depending on $s$, $D$ and $G$ values, one or three solutions of the self-consistency equation (\ref{self_consistency_m}) can be found. If the solution of (\ref{self_consistency_m}) is unique, then it is considered as the steady state solution. If three solutions are obtained, we
can assume that the lowest one corresponds to the least vegetated
stable state ($V_0$), the highest one to the most vegetated stable
state ($V_1$), while the one in the middle to the unstable state
($V_u$). If the temporal period, $T$, is sufficiently long (e.g. $T=365$ or $730$ days) to allow the dynamics
to fully develop and periodically
visit the two stable states, all the three
solutions are taken into account as steady state solutions.
If, instead, the period $T$ is much shorter (e.g. $T=182.5$ days), the dynamics are able to completely reach
only one of the two stable states, while the other one remains
unexplored. As a result, the system oscillates in time and fully
captures only one stable state, while the other one is not
approached. Only the explored stable state and the
unstable state are considered as steady state solutions.

Once solutions, $\langle V \rangle_n=\mu_n$, of
(\ref{self_consistency_m})  are determined, their conditional
probabilities, $p_n^{st}(V|G;\langle V \rangle_n)$, are computed
as

\begin{equation}
p_n^{st}(V|G;\langle V \rangle_n)=\frac 1 Z_n \textmd{exp} [(- U(V) - 2 D V^2 + 4 D \mu_n V)/s]
\end{equation}

\noindent and

\begin{equation}
Z_n=\int_{0}^{1} \textmd{exp} [(- U(V) - 2 D V^2 + 4 D \mu_n V)/s] \textmd{d}V,
\end{equation}

\noindent where $U(V)$ is the potential of the temporal
deterministic model (\ref{deterministic_model}) with water table
depth, $d$, modulated as in (\ref{d_stochastic}). Since the pdf,
$p(G)$, of the cosine function, $G =
\textmd{cos}\left(\frac{\omega \tau}{\alpha}\right)$, is

\begin{equation}
p(G)=\frac{1}{\pi \sqrt{1-G^2}},
\end{equation}

\noindent the probability $p^{st}(V; \langle V \rangle)$ can be
evaluated through the convolution integral

\begin{equation}
p^{st}(V; \langle V \rangle)=\displaystyle\int_{-1}^{+1} \left[\sum_n p_n^{st}(V|G;\langle V \rangle_n) \right] p(G) \textmd{d}G.
\end{equation}

\noindent If, for a given  value of $G$, two or more conditional
probabilities, $p_n^{st}(V|G;\langle V \rangle_n)$, exist, they are summed together and then
weighted with the same probability (defined by $p(G)$) to compute
the probability $p^{st}(V; \langle V \rangle)$. The probability
$p^{st}(V; \langle V \rangle)$ is in the end normalized so that
$\int_0^1 p^{st}(V; \langle V \rangle) \textmd{d} V=1$.

\bibliographystyle{elsarticle-harv}

\bibliography{ecol_complex_2011}

\end{document}